\documentclass[final,1p,times]{elsarticle}

\usepackage{graphics}
\usepackage{graphicx}
\usepackage{epsfig}
\usepackage{threeparttable}
\usepackage{amssymb}
\usepackage{amsthm}
\usepackage{amsmath}
\usepackage[hyphens]{url}
\usepackage{adjustbox}

\usepackage{lineno}

\biboptions{square,numbers}

\journal{arXiv}

\begin{document}

\begin{frontmatter}



\title{Resonant Ionization Laser Ion Source (RILIS) Development on Lu and Pr}


\author[TRIUMF]{R. Li \corref{cor1}}
\ead{ruohong@triumf.ca}
\author[TRIUMF,Manitoba,SFU]{J. Lassen}
\author[TRIUMF,Mary]{F. Ames}
\author[TRIUMF,SFU]{P. Kunz}
\author[TRIUMF,Manitoba]{M. Mostamand}
\author[TRIUMF]{B. B. Reich}
\author[TRIUMF,Manitoba]{A. Teigelh\"ofer}
\author[TRIUMF]{H. Yan}

\cortext[cor1]{Corresponding author}

\address[TRIUMF]{TRIUMF, 4004 Wesbrook Mall, Vancouver, BC, Canada, V6T 2A3}
\address[Manitoba]{University of Manitoba, Winnipeg, MB, Canada,  R3T 2N2}
\address[SFU]{Simon Fraser University, Burnaby, BC, Canada, V5A 1S6}
\address[Mary]{Saint Mary's University, Halifax, NS, Canada, B3H 3C3}

\begin{abstract}

Resonance ionization laser ion sources are efficient and element selective ion sources, which are particularly well suited for radioactive ion beam facilities. Using TRIUMF's off-line laser ion source test stand with a system of tunable titanium sapphire (Ti:Sa) lasers, laser resonance ionization schemes for lutetium and praseodymium have been investigated with a particular interest to autoionizing states. New ionization schemes via Rydberg states and autoionizing states were found. Their investigation and comparison of ion yields at the off-line test stand will be discussed, and the data of on-line Lu delivery will be presented. 

\end{abstract}


\end{frontmatter}



\section{Introduction}\label{Introduction}

Nuclear properties of rare-earth elements are interesting for astrophysicists due to their important role in $r$-process to form the rare-earth abundance peak in solar system~[\citenum{Mumpower}]. Furthermore with the magic number $N$=82 occurring in the mass range, the lanthanide isotopes provide a rich field for nuclear scientists  to study the shell closure and the nuclear deformation. In recent years the application of lanthanides as radioactive imaging tracer and therapy treatment for tumors increases the attention on the scientific studies on these species. However as a group of neighboring elements with similar chemical properties, it is very difficult to obtain carrier free lanthanide isotopes in high purity without contamination from the adjacent bulk of other lanthanide isotopes~[\citenum{Nayak99}]. Similar problems hampered the nuclear study on lanthanide radioactive isotopes in RIB facilities due to surface ionized isobar combinations. Based on the electronic structure of different atomic species, the Resonant Ionization Laser Ion Source (RILIS) can selectively ionize the element of interest. With the aid of a high resolution mass separator, it inherently suppresses isobaric contaminations by relatively enhancing the ionization efficiency of the aimed element. However due to the low ionization potential of lanthanides (5.4-6.4~eV), the enhancement by the RILIS is not significant over surface ionization. Recently, the introduction of the ion guide laser ion source (IGLIS) at TRIUMF~[\citenum{Seb14}] and the laser ion source trap (LIST) at CERN~[\citenum{Fink15}] opened up the possibility of isobar-free RIB of lathanides, and motivated the development of efficient laser ionization schemes on them.

Lanthanides with an open 4$f$ shell have complex electronic structures, which makes the theoretical calculation difficult and the experimental data scarce and fragmentary. However lutetium (Lu) as the last element in the lanthanide group is an exception. It has only one valence electron outside of filled 4$f$ and 6$s$ shells, which makes its atomic spectra outstandingly simple compared to other lanthanides. A series of works have done on Lu atomic structures since 1930, from traditional emission and absorption spectroscopy~[\citenum{Meg30}-\citenum{Cam72}] to contemporary laser spectroscopy~[\citenum{Maeda89}-\citenum{Vid92}]. The spectroscopic work extended to autoionizing states (AI) afterward~[\citenum{Miller82}-\citenum{Ogawa99}]. The majority of the studies on high-lying Rydberg states and AI states were made via laser resonance ionization spectroscopy (RIS), which gives helpful reference to our development work. In fact some RILIS schemes had been developed~[\citenum{Bekov81}] in early days to apply to the determination of traces of Lu in a geological study~[\citenum{Krustev93}]. The long-lived $^{176}$Lu-$^{176}$Hf radioactive decay system is a powerful tracer used in geology and astrophysics to date rocks and meteorite samples. With the high selectivity and sensitivity of the RIS method, the minimum detectable concentration can be down to 10$^{-10}$~$\%$ [\citenum{Krustev93}]. This is an another strong motivation to develop efficient schemes for Lu. The scheme used in [\citenum{Krustev93}] is 451.982~nm + 460.688~nm via an AI state at 43831.6~cm$^{-1}$ (all the wavelengths shown in this paper are the values in vacuum). In this work different atomic transitions of Lu for the first excited step (FES) were attempted. The ion yield of our new developed schemes were compared with that of the scheme in [\citenum{Krustev93}]. 

Contrary to Lu, praseodymium (Pr) has a very complicated atomic structure with three 4$f$ and two 6$s$ outer electrons. The five valence electrons give a large number of electronic configurations, very rich fine structures and strong overlap of configurations with different parities. All of these complexities resulted in the limited knowledge on spectral lines and ambiguous assignments~[\citenum{NIST}]. Although the study of Pr atomic spectra started in 1940's~[\citenum{Meggers42}], the bulk of the available data is still from the investigation before the 80's~[\citenum{Martin78}]. Even till today the knowledge of atomic properties of high-lying Rydberg states and AI states are nearly unknown. The only one reported AI state study on Pr was made using RIS by O. Kujirai \textit{et al.} in 2000~[\citenum{Kujirai00}]. On the other side, the atomic properties of Pr transitions are of increasing interest due to new astronomical observations with high-resolution spectrograph and a large number of rare earth lines observed in chemically peculiar stars~[\citenum{Ivarsson01}]. Measurements on odd-mass Pr isotopes indicate a shape transition as the neutron number increases from the magic number 82 to 90. This isotope chain provides a good candidate to test various nuclear models~[\citenum{Iimura95}]. The measurement of the nuclear moment of $^{140}$Pr was recently motived to explain the counter-intuitive observation on the decay constant of the hydrogen-like $^{140}$Pr$^{58+}$ ions in the ESR storage ring~[\citenum{Froemmgen13}]. The increasing needs for Pr isotope beams and the lack of atomic data reinforced the needs for the laser scheme development. Recently, Mainz University LARISSA group performed a laser ionization scheme development of Pr with a Ti:Sa laser system~[\citenum{Wendt14}]. The resulting optimal scheme was 461.901~nm + 899.981~nm + 877.918~nm, which ionized Pr atoms via an AI state. Meanwhile, the RILIS group at ISOLDE established a scheme of 461.901~nm + 899.996~nm + 532~nm(Nd:YAG) as a efficient scheme for their dye laser system~[\citenum{Fedosseev12}]. In this work, two different atomic transitions of Pr for the FES were tested. Especially, previously unexplored blue+blue schemes to Rydberg and AI states were developed.

\section{Experimental Setup of Offline Developments}\label{Setup}

\begin{figure}[h]
\begin{center}
\includegraphics[width=0.6 \textwidth]{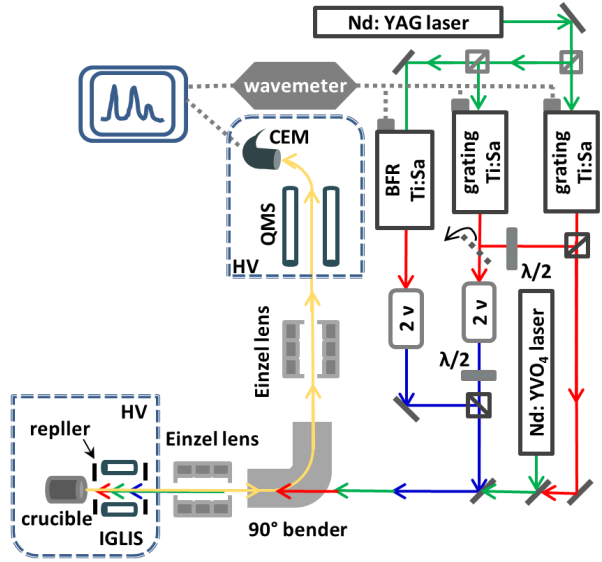}
\end{center}
\centering
\caption{The TRILIS off-line experimental setup consists of a laser system for resonant ionization, a IGLIS source~[\citenum{Seb14}], a quadrupole mass spectrometer (QMS)equipped with a channel electron multiplier (CEM), and a computer based data acquisition system.}
\label{setup}
\end{figure}

The schematic setup of the experiment is shown in Fig.~\ref{setup}. Three Ti:Sa lasers were used in the experiments, one birefringent-filter (BRF) laser and two grating lasers. All three lasers are simultaneously pumped by a pulsed Nd:YAG laser (LEE laser LDP-100MQG) of 35~W power and 10~kHz repetition rate. The Ti:Sa lasers can provide 1-2~W output power with the tunable wavelength range of 700-980~nm. The BRF laser and one of the grating lasers are both followed by a frequency doubling unit, which can extend the laser wavelength range to 350-460~nm. The typical conversion efficiency of frequency doubling is 30$\%$. The doubling unit following the grating laser was automatically controlled by a PC to phase-match the angle of the nonlinear doubling crystal (BBO or BIBO crystal) with the wavelength of the Ti:Sa laser. The wavelength of the lasers were measured by a wavemeter (High Finesse WS/6) with the precision of 10$^{-6}$. It is routinely calibrated to a polarization stabilized HeNe laser with wavelength accuracy of 10$^{-8}$ (Melles Griot 05 STP 901/903). A Nd:YVO$_4$ laser (Spectra-Physics Inc. YHP-40) is employed for nonresonant ionization schemes. This laser has a maximum output power of $\sim$12~W at 1064~nm and $\sim$6.7~W at 532~nm, with a pulse length of $\sim$30~ns. The laser beams for consecutive excitation steps were overlapped via polarization beam splitters and dichroic mirrors. The temporal superposition of laser pulses was obtained by the intra-cavity Pockel cell Q-switches. All the laser beams were finally focused into the ion source to interact with the atomic vapor of the element of interest. The source in this experiment is a IGLIS source, which consists of a thermal-ion repeller and a radio-frequency quadrupole (RFQ) as an ion guide~[\citenum{Seb14}]. The IGLIS can be operated in two different modes: transmission mode and suppression mode. In transmission mode, the potential of the crucible is higher than those of the repeller and the RFQ. Therefore, all the ions generated in the high temperature Ta crucible - either thermal ions or laser ions - transmit. However in suppression mode, the repeller has a potential higher than the crucible, so that thermal ions are repelled and prevented from entering the RFQ. So in principle, only the laser ionized ions generated after the repeller will be guided and extracted. The extracted ions are accelerated to 10~keV, then vertically bent and decelerated into a commercial quadrupole mass spectrometer (EXTREL-QMS MAX300) for mass filtering. The QMS was equipped with a channel electron multiplier (CEM) to detect the ion signal. The setup has been described in more details in [\citenum{Lav13}, \citenum{Li13}]

\section{Experimental procedure, Results and Discussion}\label{Results}

The purpose of the off-line developments of Lu and Pr is to obtain better ionization schemes based on the Ti:Sa laser system. The direct benefit is to improve the efficiency of on-line beam delivery with TRILIS and other similar LIS facilities on these two elements. It also provides good reference schemes of Lu and Pr for any other application using laser ionization as a sensitive atom detection method, such as ultra trace analysis~[\citenum{Krustev93}, \citenum{Wendt12}]. As described in our previous work~[\citenum{Li13}], no absolute ionization efficiency measurement is made in our off-line laser ionization scheme development due to the absence of a properly suppressed Faraday cup after the QMS. Thanks to the low ionization potentials of lanthanides, the surface ion yield can be used to normalize the ion yield of different laser ionization schemes, and therefore, makes the efficiencies of the these schemes comparable.  Here we use the term - relative ionization efficiency - as the ratio of laser ion yield to surface ion yield. To make sure the surface ion yield to properly act as an normalizer, a constant temperature of the source must be maintained. Furthermore the spatial laser overlapping and pointing were carefully optimized for each scheme. 

\subsection{Off-line Relative efficiency comparison of lutetium RILIS schemes}\label{Lu text}

A standard solution (Alfa Aesar Specpure, 1~$\mu$$g$/$\mu$$l$ Lu$_2$O$_3$ in 5\% HNO$_3$ solution) was loaded on a thin Zr foil. After drying out the water in a 110~$^{\circ}$C oven, the foil was folded into a tiny piece and inserted to a Ta crucible. Lu atomic vapor was generated as the crucible being resistively heated. To keep the surface ionization efficiency constant during the experiment, the temperature of the crucible was kept at 1555~$^{\circ}$C. The crucible temperature was estimated using the heating current via a calibrated temperature-current curve. Two types of temperature measurement devices, a Pyrometer (Mikron M90-R-2) and a two-color-ratio fiber-optic-infrared temperature-measurement system (Omega iR2C), had been used for calibration. The measurement precision is within 1.5$\%$, and the measurement accuracy of two different devices is within 15$\%$.

\begin{figure}[h]
\begin{center}
\includegraphics[width=0.6\textwidth]{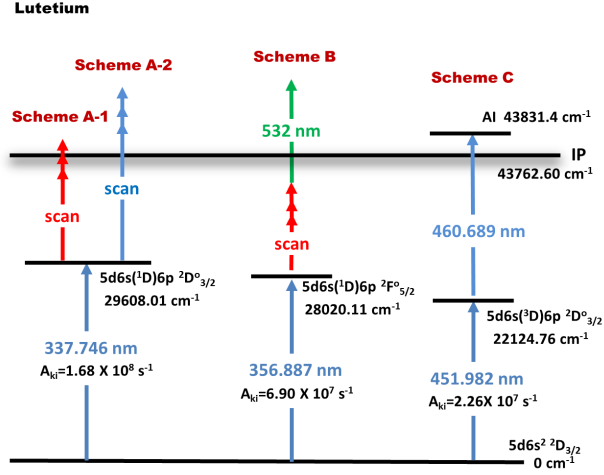}
\end{center}
\centering
\caption{Lutetium ionization schemes. A$_{ki}$ is the Einstein A coefficient. The FES were all provided by the frequency doubled BRF laser. The SES and TES were provided by the grating lasers. The red colored line means the fundamental output of the laser was applied. The blue colored line means the frequency doubled output of the used laser. Scheme C is from [\citenum{Krustev93}] and has been used at ISAC for Lu yield measurement (Sect.~[\ref{Lu on-line text}].}
\label{Lu_scheme}
\end{figure}

\begin{figure*}[!htbp]
\begin{center}
\includegraphics[width=1\textwidth]{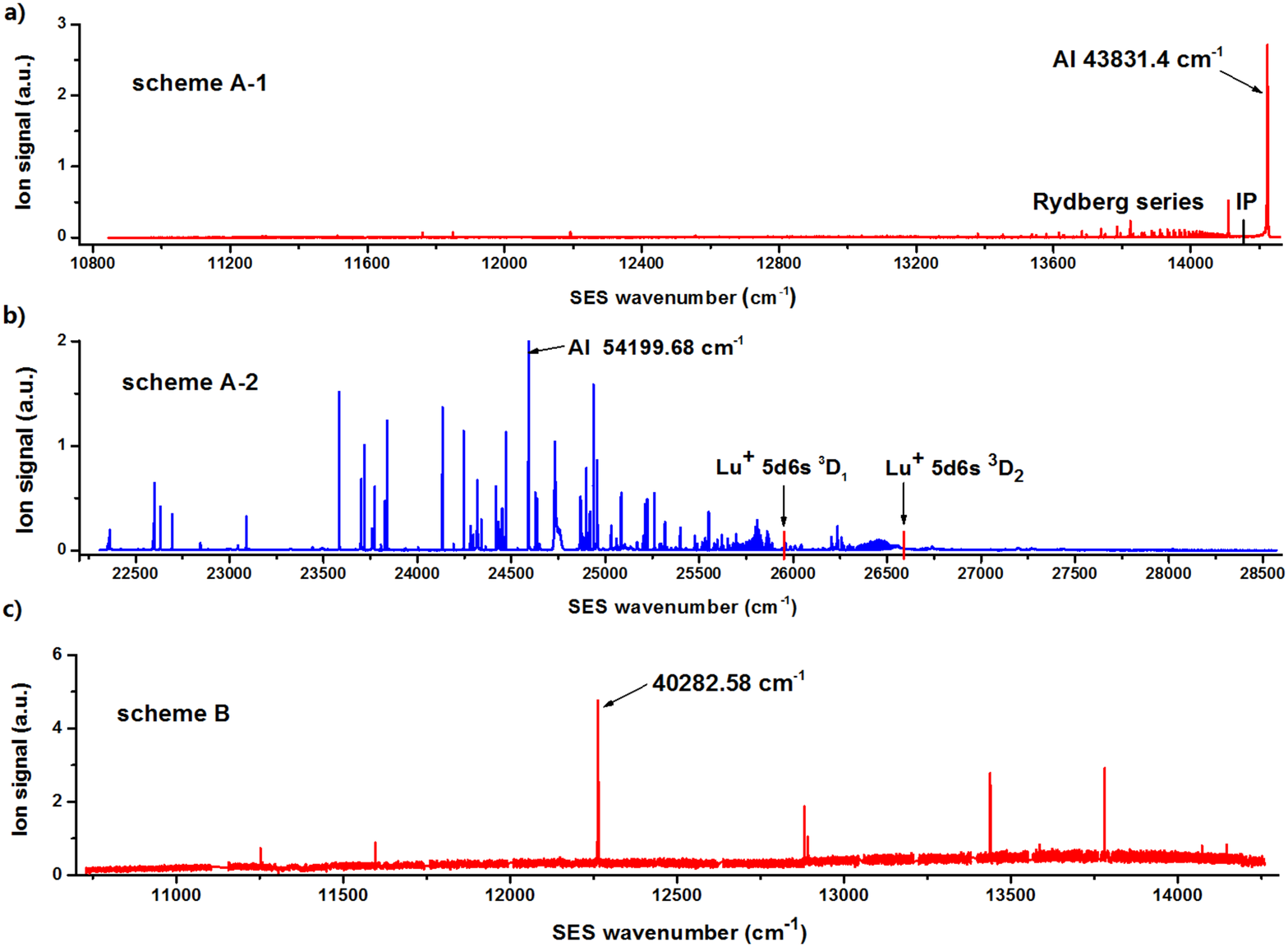}
\end{center}
\centering
\caption{Lutetium spectra: a) Scheme A-1 spectrum: three Rydberg series approaching the IP were observed. The AI state at 43831.6~cm$^{-1}$ provides the best ion yield in this spectrum. b) Scheme A-2 spectrum: AI Rydberg series approaching ionic states Lu$^+$ 5d6s $^3$D$_1$ and Lu$^+$ 5d6s $^3$D$_2$ were observed. The AI state at 54199.68~cm$^{-1}$ gives the best ion yield in this spectrum. c) Scheme B spectrum: eleven excited states were observed. The state at 40282.58~cm$^{-1}$ gives the best ion yield in this spectrum.}
\label{Lu_spectra}
\end{figure*}

The investigated ionization schemes of Lu are shown in Fig.~\ref{Lu_scheme}. In both, scheme A-1 and scheme A-2, Lu atoms were firstly excited from the ground state 5d6s$^2$ $^2$D$_{3/2}$ to an intermediate state of 5d6s6p $^2$D$^{\circ}_{3/2}$. The laser of this excitation step was provided by the frequency-doubled BRF laser. The laser for the second excitation step (SES) was the fundamental output or the frequency doubled output of the grating laser, for scheme A-1 and scheme A-2 respectively. The fundamental output of the grating laser has a continuously tunable wavelength range of 720-930~nm~[\citenum{Tei10}], which can excite the Lu atoms from the intermediate state into various even-parity Rydberg states. The obtained spectrum is shown in Fig.~\ref{Lu_spectra}-a. Three series of Rydberg states 6s$^2$nd $^2$D$_{3/2}$, 6s$^2$nd $^2$D$_{5/2}$ and 6s$^2$ns $^2$S$_{1/2}$ were observed with clear regularity approaching the ionization potential (IP). The spectroscopic analysis and discussion of the Rydberg states is presented in [\citenum{LiLu16}]. In the obtained spectrum of this scheme, the highest ion yield was achieved by exciting the Lu atoms to an AI state at 43831.6~cm$^{-1}$. This AI transition has a linewidth of 2.5~cm$^{-1}$, which also gives a good stability against small frequency drifts during on-line RIS beam delivery. It is worth noting that this AI state is the same one used in [\citenum{Krustev93}], the scheme of which is marked as scheme C in Fig.~\ref{Lu_scheme}.

In scheme A-2, the automatic phase-matching of the nonlinear crystal was employed, which enables a continuous wavelength scan of the frequency doubled output of the grating Ti:Sa laser. Due to the angle change of the nonlinear crystal during the wavelength scan and the loose restriction of the laser cavity in the grating incident plane, a regular manual correction of the laser spatial pointing was needed. The necessity of correction is referred to the significantly drop on the resonance signal and the visual offset of laser reflections on a reference screen. Normally the correction of the alignment was made by optimizing on a resonance signal every 10-15~nm. This resonance signal was used as a normalization to link the scans. The resulting spectrum is shown in Fig.~\ref{Lu_spectra}-b. The scan covers the energy range of 51960 - 58008~cm$^{-1}$, across two ionic state limits 5d6s $^3$D$_1$ and 5d6s $^3$D$_2$. The detailed analysis of atomic spectroscopy is presented in [\citenum{LiLu16}]. The output power of the frequency doubled laser varied from 40~mW at two ends of 370~nm and 445~nm, and 200~mW at the middle wavelength of 420~nm. Due to the low output at the ends of the tuning curve, the effect of the laser power on the ion yield may be not negligible. Based on this condition, the best scheme in the spectrum of Fig.~\ref{Lu_spectra}-b is the one via the AI state at 54199.68~cm$^{-1}$ with the linewidth of 0.4~cm$^{-1}$.

Scheme B is a three step scheme with nonresonant ionization. The FES excites Lu atoms from the ground state to the 5d6s6p $^2$F$^{\circ}_{5/2}$ state at 28020.11~cm$^{-1}$. The TES, namely the nonresonant ionization step, was provided by the 4.8~W Nd:YVO$_4$ at 532~nm. To search the best SES for this scheme, a wavelength scan of the fundamental output of the grating laser was performed. The resulting spectrum is shown in Fig.~\ref{Lu_spectra}-c. The highest ion yield in the spectrum is given via an excited state at 40282.58~cm$^{-1}$, whose electronic configuration has not been clearly assigned. Scheme C had been developed and investigated in [\citenum{Xu93}-\citenum{Kujirai98}], and had been practically applied to ultratrace measurements in~[\citenum{Krustev93}]. Measurement of this scheme was also performed in our experiments to compare its ionization efficiency with that of the new schemes. The FES of scheme C is from the ground state to 5d6s6p $^2$D$^{\circ}_{3/2}$ state at 22124.76~cm$^{-1}$. And the SES is exciting Lu atoms further to the AI state at 43831.6~cm$^{-1}$.

\begin{table*} [!htbp] \footnotesize
\caption{Lutetium: the relative ionization efficiency $\epsilon$$_{rel}$ comparison of the different schemes (Fig.~\ref{Lu_scheme}). $P_{sat}$ is the saturation power of the transition, and $P_{op}$ is the laser power in operation when measuring the scheme efficiency. The $\lambda$$_1$ for the FES are from NIST ASD~[\citenum{NIST}], and the $\lambda$$_2$ for the SES are from our measurements. The errors of the $P_{op}$ are about $\pm$2 in the last digit. For all the measurements, the temperature of the crucible was kept at 1555~$^{\circ}$C, corresponding to an crucible heating current of 180~A. The power of the 532~nm laser was 4.8~W.}
\begin{center}
\begin{threeparttable}
\begin{tabular}{ccccccccc}
	\hline\hline
\\
&\multicolumn{2}{c}{FES}&&\multicolumn{2}{c}{SES}&&&\\
\cline{2-3} \cline{5-6}
scheme&$\lambda$$_1$&$P_{op}$/$P_{sat}$&&$\lambda$$_2$&$P_{op}$/$P_{sat}$&ionization&$\epsilon$$_{rel}$&$\epsilon$$_{rel}$\\
 &(nm)&(mW)&&(nm)&(mW)&via&&+532nm\\
\hline
A-1&337.746&30~/~6(1)&&703.056&220~/~95(7)&AI\tnote{a}&3.0&7.5\\
A-2&337.746&30~/~6(1)&&406.642&79~/~h\tnote{b}&AI&2.0&5.5\\
B&356.887&110~/~1.9(3)&&815.497&1300~/~61(8)&Non&1.0&10\\
C\tnote{c}&451.982&145~/~8(2)&&460.684&45~/~27(2)&AI\tnote{a}&1.5&7.0\\
\hline\hline

\end{tabular}
\begin{tablenotes}
       \item[a] Two schemes use the same AI state at 43831.6(3)~cm$^{-1}$.
       \item[b] The saturation power is too high to be accurately extracted from the measured saturation curves.
       \item[c] The scheme has been investigated~[\citenum{Xu93}-\citenum{Kujirai98}] and applied to ultratrace measurement~[\citenum{Krustev93}]. On-line isotope yield using this scheme at ISAC is presented in Sect.~\ref{Lu on-line text}.
\end{tablenotes}
\end{threeparttable}
\end{center}
\label{table_Lu_EF}
\end{table*}

To determine the saturation power of the transitions, the ion yield dependences on the laser powers were measured. For convenience, the laser power was measured on the laser table right after the telescope expansion. After that, the laser beam will still lose power by passing the combination mirrors, transportation prism and the vacuum window before it finally interacts with the Lu atoms. A typical transportation loss of $\sim$50$\%$ shall be taken into account. The exciting and ionizing lasers all overlap in the 3~mm diameter crucible. In the laser alignment prior to the experiment, the laser beams were focused as small as possible into the crucible region. During the experiment, a slight readjustment was attempted to maximize the ion yield. In principle, the maximum signal is achieved when the laser beams spatially match with the atomic vapor volume (3~mm diameter) at the saturated power. A function of $I=I_0+A(1-e^{-P/P_{sat}}$) was applied to fit the measured data, where $I$ is the ion yield, $P$ is the laser power and $P_{sat}$ is the saturation power. In some cases, Lu atoms may absorb one more photon from the resonate laser after being excited to the aimed state, and ionize nonresonantly if the total energy is over the IP. In these cases, a nonresonant term $A'*P$ was added to the saturation function above. The measured saturation powers for the tested excitation schemes are listed in Tab.~\ref{table_Lu_EF}.

Tab.~\ref{table_Lu_EF} also presents the measured relative laser ionization efficiency $\epsilon$$_{rel}$ of the schemes. The $\epsilon$$_{rel}$ was defined as the enhancement ratio of laser ionization over surface ionization. For easy comparison, the values were normalized to the $\epsilon$$_{rel}$ of scheme B. The laser powers used in measuring the efficiencies are listed as $P_{op}$. Comparing the $P_{op}$ with the $P_{sat}$ reveals the possibility to improve the $\epsilon$$_{rel}$ by increasing the laser power. The 532~nm laser pronouncedly increased the $\epsilon$$_{rel}$. With only the 532~nm laser into the source, the surface ionization of the Lu did not increase evidently, which seemed to imply the atom vapor density was not significantly boosted by the laser. Therefore, it was not a dominant contributor to the enhancement of $\epsilon$$_{rel}$. However, when adding the 532~nm laser to the AI scheme A-1, there were two distinguishable effects: one was immediately seen right after adding the laser, and the other one ramped up gradually in $\sim$1 minute. Delaying the pulse of the 532~nm laser off the synchronization around more than 3~$\mu$s, both effects persisted. A similar phenomenon was also observed in Pr development, which is discussed later in this paper.

\subsection{On-line lutetium RILIS test}\label{Lu on-line text}

\begin{table} [!htbp]\footnotesize
\caption{Lutetium: on-line ion yield of an IGLIS source in transmission mode. The isotopes were extracted from a UCx target bombarded with 10~$\mu$A 480~MeV proton beam. The ion yield was measured as the count difference on a channeltron between laser on and off.}
\begin{center}
\begin{threeparttable}
\begin{tabular}{cccccc}
\hline\hline
isotope	&	half-life	&	target T	&	lasers	&	lasers	&	enhancement 	\\
	& &	&	off	&	on	&	ratio	\\
	& &	($^{\circ}$C)	&	(s$^{-1}$)& (s$^{-1}$) &	\\

\hline
\\											
$^{176}$Lu	&	3.7~h/3.8$\times$10$^{10}$~a	&	1550	&	1.2$\times$10$^{4}$	&	3.6$\times$10$^{5}$	&	30	\\
$^{177}$Lu	&	7~m/166~d/6.6~d	&	1550	&	5.0$\times$10$^{3}$	&	1.4$\times$10$^{5}$	&	28	\\
$^{177}$Lu	&	7~m/166~d/6.6~d	&	1650	&	4.0$\times$10$^{4}$	&	4.0$\times$10$^{5}$	&	10	\\
$^{178}$Lu	&	22.7~m/28.4~m	&	1650	&	/	&	not seen	&	/	\\
$^{179}$Lu	&	4.6~h	&	1650	&	/	&	tiny	&	/	\\
\\
\hline\hline
\end{tabular}
\end{threeparttable}
\end{center}
\label{table_Lu_online_IGLIS}
\end{table}

We have applied the scheme C to an IGLIS source with a 10~$\mu$A proton beam hitting on a uranium carbide (UCx) target in June 2015. The generated radioactive isotopes effuse into the IGLIS source, where laser ionization occurs. The IGLIS source was set in a transmission mode~[\citenum{Seb14}]. The temperature of the UCx target was estimated by the heating current via the pre-measured temperature-current curves and the calculated power deposition from the proton beam. The yield of the Lu isotope was measured as an ion count on a downstream channeltron. The saturation powers measured on line were 10(1)~mW and 33(2)~mW for the FES and SES, respectively. The disagreement with the values obtained (Tab.~\ref{table_Lu_EF}) can be explained with the different transportation efficiency of the laser beams. The ion yields for different isotopes of Lu are shown in Tab.~\ref{table_Lu_online_IGLIS}. The laser powers are 250~mW and 600~mW for the FES and SES respectively during the yield measurement. From the result, we can see the release of Lu isotopes from UCx target is not favored, especially for short lifetime isotopes. We did not observe any Lu$^+$ release for the isotopes whose lifetime is shorter than 5 hours. Additionally, the surface ionized Lu$^+$ was barely seen. 

In June 2016, we applied the scheme C to a rhenium (Re) hot cavity RILIS source with a 60~$\mu$A proton beam on a Ta target. The release of Lu isotopes from this target is much better than from a UCx target. The temperature of the target was about 2300~$^{\circ}$C. Here the ion yield was measured with a downstream channeltron and the ISAC yield station~[\citenum{Kunz}]. The investigated isotopes yields are listed in Tab.~\ref{table_Lu_online_hot}.

\begin{table} [!htbp] \footnotesize
\caption{Lutetium: on-line ion yield from a Ta target with a Re hot cavity source bombarded with a 60~$\mu$A proton beam. Here, the ion yield was measured with the ISAC yield station through nuclear decay counting. All yields in this table was evaluated via characteristic $\gamma$ lines.}
\begin{center}
\begin{threeparttable}

\begin{tabular}{cccccc}
\hline\hline
isotope	&	isomer & half life	& laser &	laser	&	enhancement	\\
	&		&&	laser off	& laser on	&	ratio	\\
	&		&&	(s$^{-1}$)	&	(s$^{-1}$)	&		\\
\hline
\\									
$^{164}$Lu	&	g	&	3.14~m	&	2.1$\times$10$^{8}$	&	5.5$\times$10$^{8}$	&	2.6	\\
$^{166}$Lu	&	g	&	2.65~m	&	1.6$\times$10$^{8}$	&	4.4$\times$10$^{8}$	&	2.8	\\
$^{166}$Lu	&	m1	&	1.41~m	&	1.5$\times$10$^{8}$	&	1.7$\times$10$^{8}$	&	1.1	\\
$^{166}$Lu	&	m2	&	2.12~m	&	7.5$\times$10$^{7}$	&	2.0$\times$10$^{8}$	&	2.7	\\
$^{167}$Lu	&	g	&	51.5~m	&	8.6$\times$10$^{8}$	&	3.1$\times$10$^{9}$	&	3.6	\\
$^{168}$Lu	&	g	&	5.5~m	&	2.0$\times$10$^{8}$	&	6.8$\times$10$^{8}$	&	3.4	\\
$^{168}$Lu	&	m1	&	6.7~m	&	6.5$\times$10$^{8}$	&	1.7$\times$10$^{9}$	&	2.6	\\
$^{169}$Lu	&	g	&	1.42~d	&	3.1$\times$10$^{9}$	&	2.3$\times$10$^{10}$	&	7.4	\\
$^{170}$Lu	&	g	&	2.0~d	&	3.1$\times$10$^{9}$	&	1.4$\times$10$^{10}$	&	4.5	\\
$^{171}$Lu	&	g	&	8.24~d	&	2.7$\times$10$^{9}$	&	2.0$\times$10$^{10}$	&	7.4	\\
$^{172}$Lu	&	g	&	6.7~d	&	5.9$\times$10$^{8}$	&	3.9$\times$10$^{9}$	&	6.6	\\
$^{177}$Lu	&	g	&	6.647~d	&	4.1$\times$10$^{7}$	&	6.5$\times$10$^{8}$	&	16	\\
$^{178}$Lu	&	g	&	28.4~m	&	1.0$\times$10$^{7}$	&	4.2$\times$10$^{7}$	&	4.2	\\
\\
\hline\hline
\end{tabular}
\end{threeparttable}
\end{center}
\label{table_Lu_online_hot}
\end{table}

\subsection{Temperature dependence of the ion yield via Rydberg states of lutetium}\label{Lu temperature text}

With the total excitation energy lower than the IP, the detailed ionization mechanism of Rydberg states is still an open question. For Lu a high ion yield peak is generated via the Rydberg state at 43717.68~cm$^{-1}$, which is only $\sim$45~cm$^{-1}$ beneath the IP. The ion yield is outstandingly high compared to the neighboring Rydberg states, and more interestingly, it increases significantly (relative to that of the AI state at 43831.6~cm$^{-1}$) with the ambient temperature. A similar phenomenon was seen in the laser ionization spectroscopy of antimony~[\citenum{Li16}]. The occurrence of this phenomenon in both cases coincides with the existence of a perturbation from doubly excited states. The mixture with the doubly excited valence states significantly decreases the lifetime of the state, \textit{e.g.} the photoexcitation rate to the state, therefore prominently enhances the ion yield. The detail of the perturbation is discussed in [\citenum{LiLu16}]. 

\begin{figure}[!htbp]
\begin{center}
\includegraphics[width=0.8\textwidth]{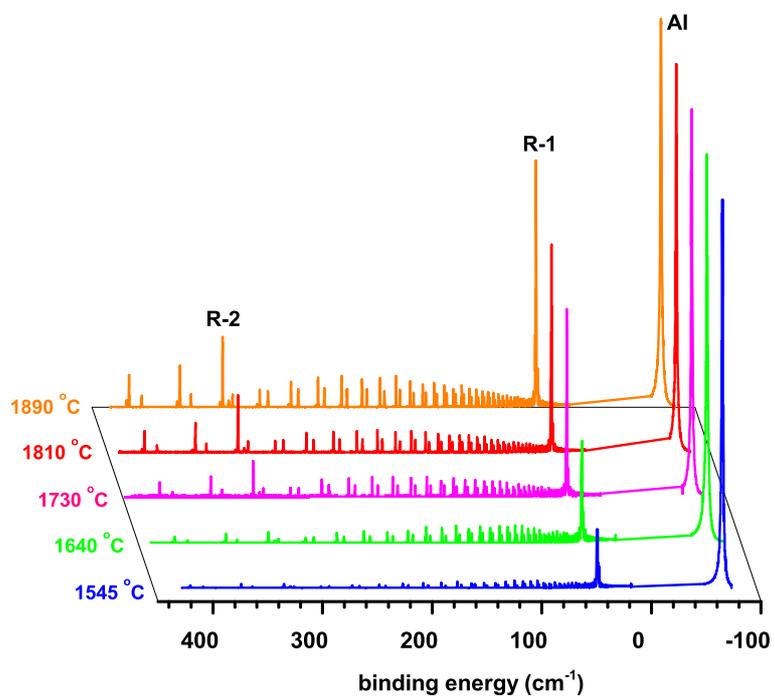}
\end{center}
\centering
\caption{Lutetium: the ionization via Rydberg states depends on the ambient temperature. The IGLIS was set in the transmission mode. The spectra at different temperatures have been normalized to the AI resonance peak.}
\label{Lu_temperature}
\end{figure}

The interesting operational aspect for the RILIS is the extraordinary enhancement of ion yield when increasing the ambient temperature. The laser ionization spectra of high Rydberg states and the AI state 43831.6~cm$^{-1}$ were obtained under different crucible temperatures. The AI resonance peak acts as a reference to normalize the Rydberg state spectra. The experimental results are shown in Fig.~\ref{Lu_temperature}. R-1 is the perturbed Rydberg state at 43717.68~cm$^{-1}$, and R-2 is another perturbed Rydberg state at 43431.82~cm$^{-1}$. Unlike the very localized line-intensity change at R-1, the perturbation around R-2 is much broader, which indicates a strong perturbation. Despite the different magnitude of perturbation, the two perturbed Rydberg states both show dependence of the ion yield on the temperature, so do some other regular Rydberg states. Due to the broad perturbation around R-2, \textit{e.g.} multiple Rydberg states are mixed with a doubly excited state. It is difficult to tell whether this temperature-dependence is strongly affected by the mixture of the doubly excited state, or only related to the binding energy. However, evident is that at higher temperature the ionization probability increases at higher binding energy range. This should either relate to the blue-shift of maximum emission of the black body radiation spectrum ($\lambda$$_{max}$$\sim$${1/T}$) with increased temperature, or be caused by the increase of the thermal kinetic energy of particles ($E$$_k$$\sim$${kT}$) in the collisions. More investigations are need to determine whether both mechanisms significantly contribute or only one dominates, and whether the leading mechanism changes when the Rydberg states mix with doubly excited states. Additionally, the total photon emission from the black body radiation increases with temperature, so does the collision rate. However this effect should equally address on all Rydberg resonances. In conclusion, when applying the laser ionization schemes with Rydberg states the temperature must be considered as an important parameter affecting the ionization efficiency.

\begin{figure}[!htbp]
\begin{center}
\includegraphics[width=0.8\textwidth]{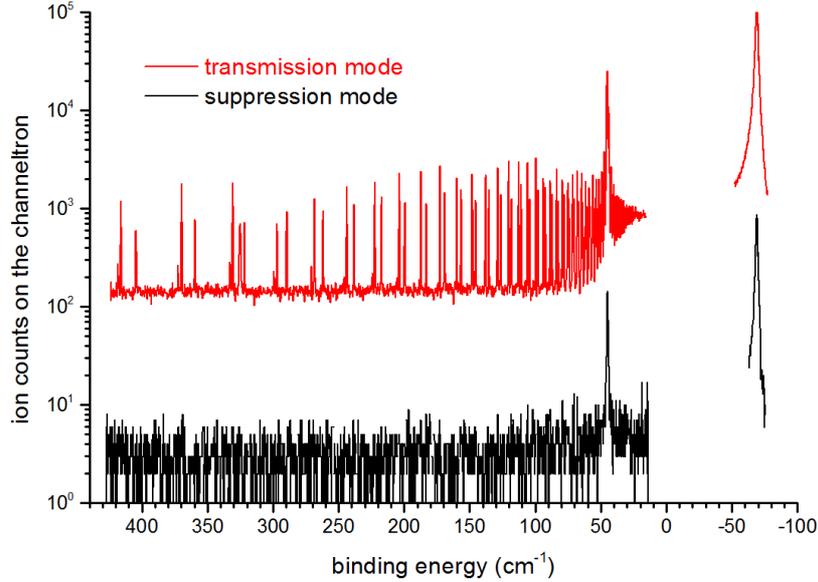}
\end{center}
\centering
\caption{Lutetium: The ionization efficiency via Rydberg state and AI state in both transmission and suppression mode of IGLIS. The result evidently showed the elimination of the Rydberg state ionization when the IGLIS was operated at the suppression mode. The binding energy range from 0~cm$^{-1}$ to 60~cm$^{-1}$ was intentionally skipped with the advance knowledge of no resonance in this region.}
\label{suppression_Rydberg}
\end{figure}

Proven the temperature effect on the ion yield of Rydberg states, the feasibility to use Rydberg-state RILIS schemes with an IGLIS was tested. The laser ionization spectra covering the high-lying Rydberg states and the 43831.6~cm$^{-1}$ AI state were measured at both transmission and suppression mode of IGLIS. To make the measurements comparable, the only difference of ion optics between the two modes is the increasing the repeller potential to +22~V in the suppression mode instead of -4~V relative to the crucible potential in the transmission mode. The crucible temperature was kept constant (1555~$^{\circ}$C), so were all the laser settings. In the transmission mode, the majority of laser ionization happens in the hot crucible. However in the suppression mode, those ions are pushed backward by the potential of the repeller and not extracted. Only a small amount of ions (around 2 orders of magnitude less shown in the Fig.~\ref{suppression_Rydberg}) generated after the repeller was guided by the RF and finally get detected. Approximately we can say the resonant ions signal observed in two modes are generated at different locations: inside the crucible and after repeller, for the transmission mode and the suppression mode, respectively. Inside the hot crucible, Rydberg atoms can easily get ionized by thermal photons and collisions. However after the repeller, both the density of thermal photons and the collision possibility drop significantly, so does the possibility of ionizing Rydberg atoms. Consequently as shown in Fig.~\ref{suppression_Rydberg}, no Rydberg resonant peaks are observed when the IGLIS operated in the suppression mode.

An interesting phenomenon is that the perturbed Rydberg state R-1, or the component from the doubly-excited valence state, still ionizes at the IGLIS suppression mode. Assuming that the neutral atoms effuse from the 3 mm diameter hot crucible and form a beam with an opening angle $\sim$10$^{\circ}$, the relevant ionization volume in suppression mode should locate directly downstream from the repeller electrode. However this region is not field-free. In this experiment, the repeller has a potential +20~V relative to the RFQ. And the RFQ has a $\sim$75~V$_{pp}$ 0.3~MHz square-wave alternative potential on the rods. According to the classical saddle point model~[\citenum{Bransden83}], the ionization threshold W$_{th}$(F) is lowered by the electric field:
\begin{equation}
W_{th}(F)=IP-2\sqrt{\dfrac{Z_{eff}e^{3}}{4\pi\epsilon_0}}\sqrt{F},
\end{equation}
where $IP$ is the ionization potential in zero field, Z$_{eff}$ is the effective charge number of the core ($\approx$1 for high Rydberg states), $e$ is the electric charge of the electron and $\epsilon$$_0$ is the permittivity of the vacuum. From the formula, to shift the ionization threshold 50~cm$^{-1}$ below the IP, 75~V/cm is needed. This field strength was proven to reasonably exist in the suppression mode by SIMION simulation~[\citenum{SIMION}]. With the lowering of the ionization threshold, the doubly excited valence state at R-1 will be forced to become an AI state. This phenomenon of \textquotedblleft forced autoionization" was observed at ~200~cm$^{-1}$ below the IP of barium and was theoretically explained by Sandner \textit{et al.}~[\citenum{Sandner86}]. With the external electrical filed strength of 133~V/cm, a forced AI of Lu was observed by Maeda \textit{et al.}, which is exactly at the same energy position of R-1~[\citenum{Maeda89}]. The forced ionization spectrum they obtained agrees with our observation (Fig.~\ref{suppression_Rydberg}). The onset of stark continuum at W$_{th}$(F) was not evident in our spectrum due to the field inhomogeneity within the ionization volume.

\begin{figure*}[!htbp]
\begin{center}
\includegraphics[width=1\textwidth]{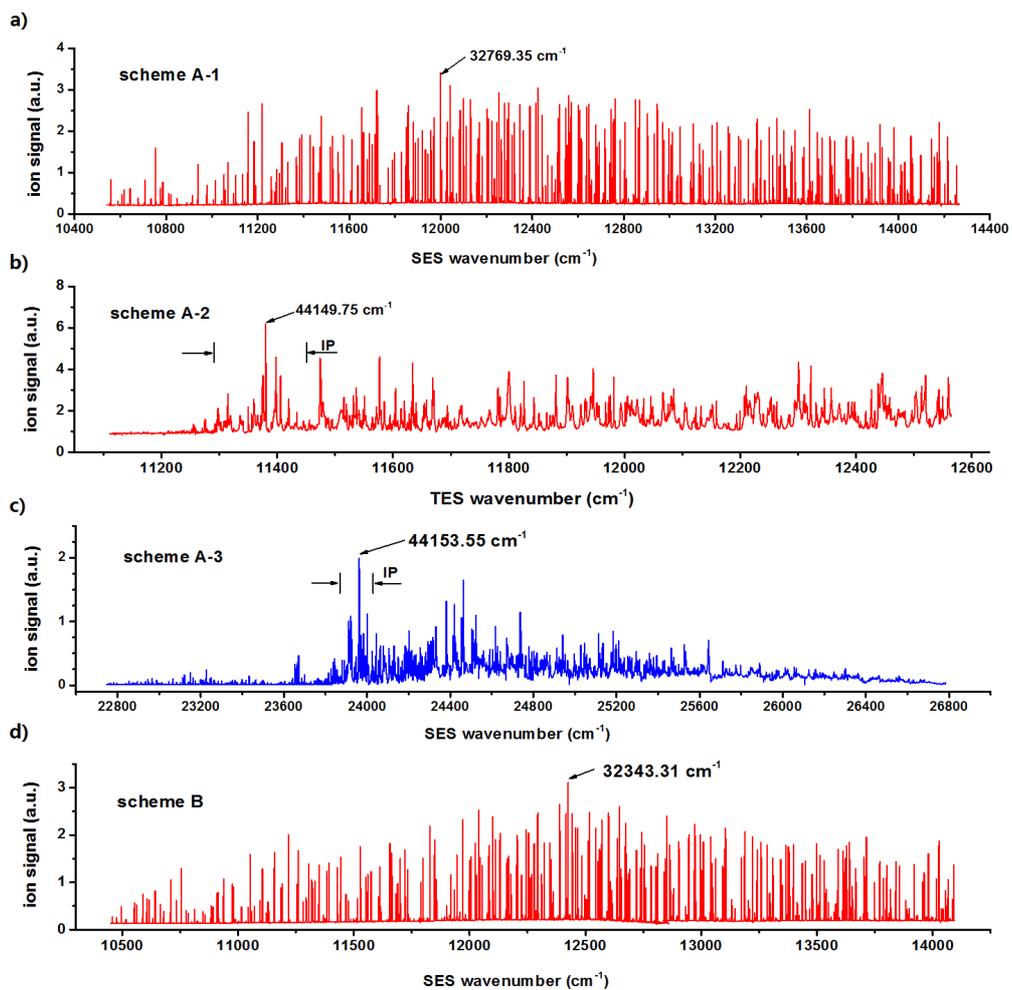}
\end{center}
\centering
\caption{Praseodymium spectra: a) Scheme A-1 spectrum: the intermediate state at 32769.35~cm$^{-1}$ gives best ion yield in this spectrum. b) Scheme A-2 spectrum: the Rydberg/AI state at 44149.54~cm$^{-1}$ gives best ion yield in this spectrum. The IP value with uncertainty~[\citenum{NIST}] is marked. c) Scheme A-3 spectrum: the Rydberg/AI state at 44153.03~cm$^{-1}$ gives best ion yield in this spectrum. d) Scheme B spectrum: the intermediate state at 32343.04~cm$^{-1}$ gives best ion yield in this spectrum. }
\label{Pr_spectra}
\end{figure*}

\subsection{Off-line Relative efficiency comparison of praseodymium RILIS schemes}

\begin{figure}[!htbp]
\begin{center}
\includegraphics[width=0.6\textwidth]{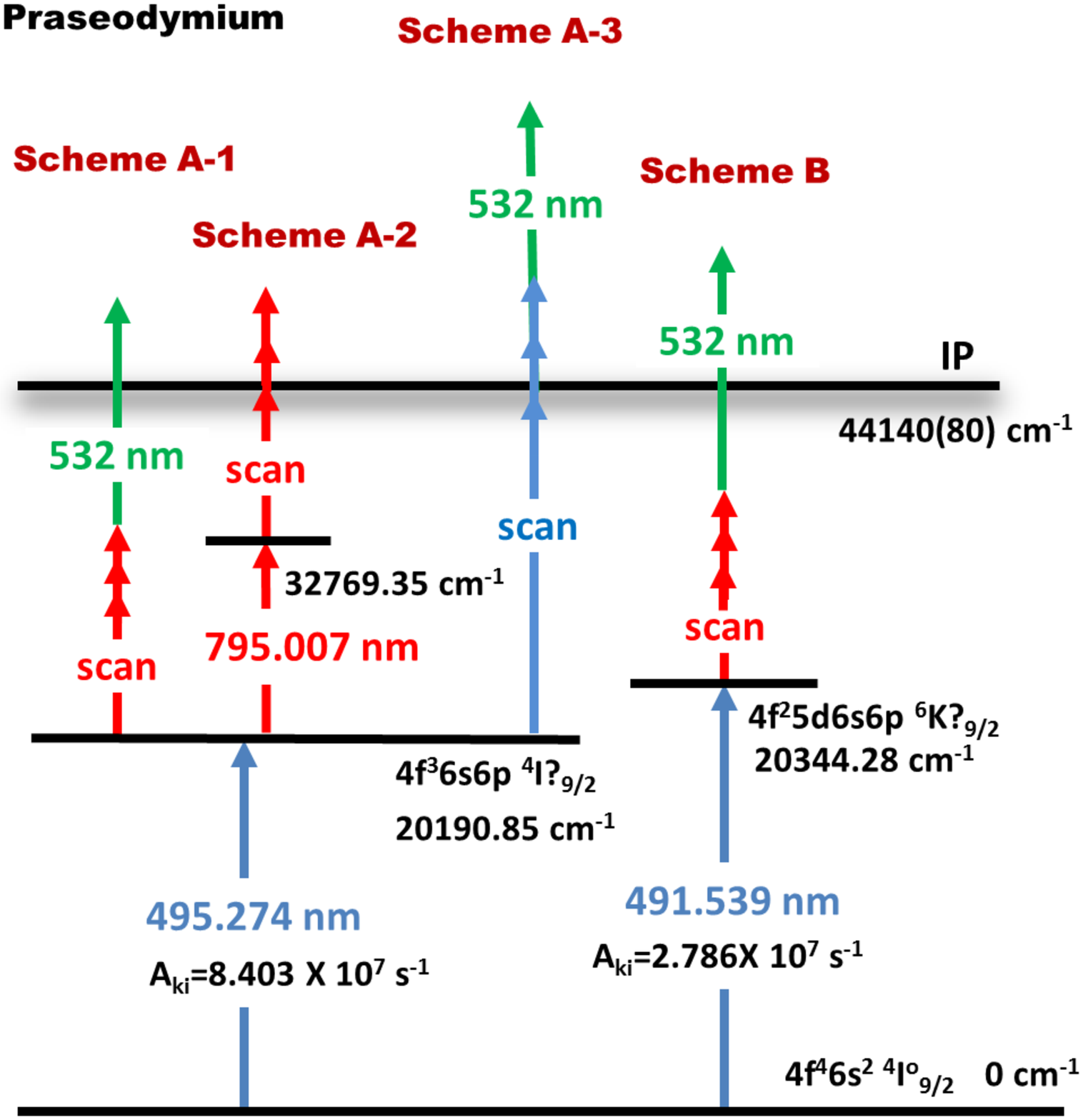}
\end{center}
\centering
\caption{Praseodymium ionization schemes. A$_{ki}$ is the Einstein A coefficient. The meanings of the dash line, solid line, line colors are the same as Fig.~\ref{Lu_scheme}.}
\label{Pr_scheme}
\end{figure}

A standard solution of Alfa Aesar Specpure (1~$\mu$$g$/$\mu$$l$ Pr$_6$O$_{11}$ in 5\% HNO$_3$) was deposited on a titanium (Ti) foil and then heated inside a Ta crucible. For the whole experiment, the crucible was kept in a same temperature of 1555~$^{\circ}$C, corresponding to the crucible heating current of 180~A. The Pr laser excitation schemes studied in this work are shown in Fig.~\ref{Pr_scheme}. For all the schemes A (A-1, A-2 and A-3), the Pr atoms were firstly excited into an intermediate state 4f$^3$6s6p $^4$I?$_{9/2}$~[\citenum{NIST}] at 20190.85~cm$^{-1}$. From this state the excitation schemes using infrared (IR)+532~nm, IR+IR and blue+532~nm were investigated. The power of the 532~nm laser was 6.7~W. The experimental techniques of continuous infrared and blue scans were the same as described in Sect.~\ref{Lu text}. The resulting spectra are shown in Fig.~\ref{Pr_spectra}.

The transitions that generated highest ion yield in each spectrum are marked in Fig.~\ref{Pr_spectra}, and listed in Tab.~\ref{table_Pr_EF} to compare the relative ionization efficiencies $\epsilon$$_{rel}$. The values of $\epsilon$$_{rel}$ were normalized to that of scheme A-1. In Fig.~\ref{Pr_spectra} b and c, the upper states, giving best ion yield, are within the uncertainty of the best known IP value, therefore they could be Rydberg or AI states. The SES of scheme A-2 was chosen as 795.007~nm transition because this transition as the SES in scheme A-1 gives the highest ion yield. The corresponding upper state is 32769.35~cm$^{-1}$ (Fig.~\ref{Pr_spectra}-a). The spectral linewidth of the all the transitions for the best ion yield (including the two Rydberg/AI states) is 9-15~GHz, dominated by the laser linewidth. 

\begin{figure}[!htbp]
\begin{center}
\includegraphics[width=0.8\textwidth]{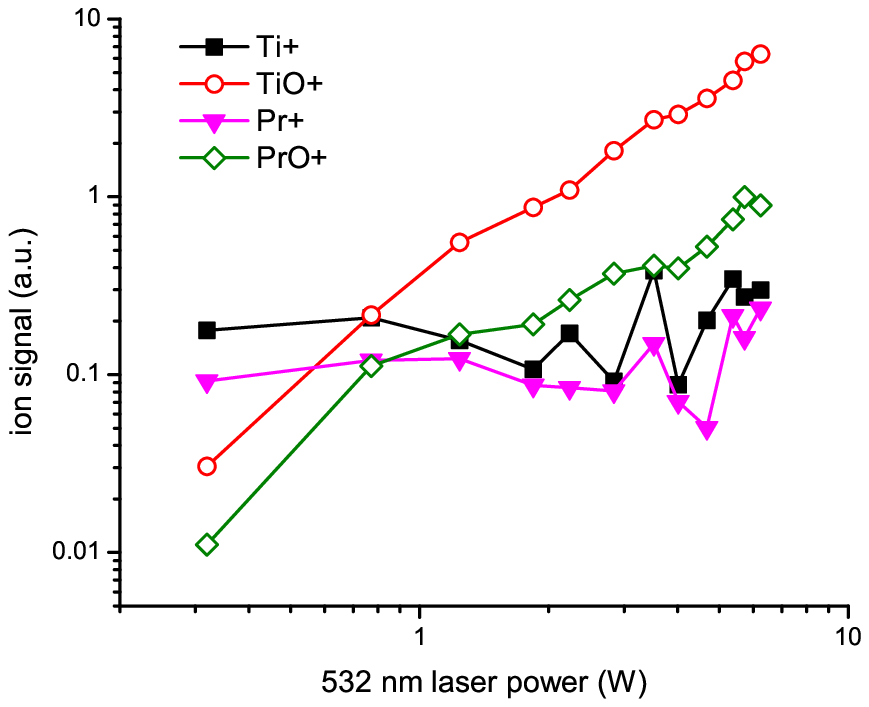}
\end{center}
\centering
\caption{Molecular and atomic ion yield dependence on the 532~nm laser power. Pr$^+$ was generated by scheme A-3 and both steps were saturated. Ti$^+$ was surface ionized. }
\label{molecular}
\end{figure}

\begin{table*} [!htbp] \footnotesize
\caption{Praseodymium: the relative ionization efficiency $\epsilon$$_{rel}$ comparison of the different schemes. The scheme details are shown in Fig.~\ref{Pr_scheme}. the meanings of $P_{sat}$, $P_{op}$ and $\lambda$ are the same as what described in Tab.~\ref{table_Lu_EF}. The $\lambda$$_1$ for the FES are from NIST ASD~[\citenum{NIST}], and the $\lambda$$_2$ for the SES and the $\lambda$$_3$ for the TES are from our measurements. The errors of the $P_{op}$ are about $\pm$2 in the last digit. For all the measurements, the temperature of the crucible was kept at 1555~$^{\circ}$C, corresponding to an crucible heating current of 180~A. The power of the 532~nm laser was 6.7~W.}
\begin{center}
\begin{threeparttable}
\begin{tabular}{cccccccccccc}
	\hline\hline
\\
&\multicolumn{2}{c}{FES}&&\multicolumn{2}{c}{SES}&&\multicolumn{2}{c}{TES}&&&\\
\cline{2-3} \cline{5-6} \cline{8-9}
scheme&$\lambda$$_1$&$P_{op}$/$P_{sat}$&&$\lambda$$_2$&$P_{op}$/$P_{sat}$&&$\lambda$$_3$&$P_{op}$/$P_{sat}$&ionize&$\epsilon$$_{rel}$&$\epsilon$$_{rel}$\\
 &(nm)&(mW)&&(nm)&(mW)&&(nm)&(mW)&via&&+532nm\\
\hline
A-1&495.274&390/6(1)&&795.007& 1130/460(90)&&-&-&Non&1.0&6.0\\
A-2&495.274& 390/6(1)&&795.007& 1130/460(90)&&878.704&480/12(1)&Ryd/AI&5.2&12\\
A-3&495.274&390/19(3)&&417.315&180/h\tnote{a}&&-&-&Ryd/AI&2.1&4.3\\
B&491.539& 515/7(2)&&833.401& 1370/36(7)&&-&-&Non&4.3&13\\
\hline\hline

\end{tabular}
\begin{tablenotes}
       \item[a] The saturation power is too high to be accurately extracted from the measured saturation curves. 
\end{tablenotes}
\end{threeparttable}
\end{center}
\label{table_Pr_EF}
\end{table*}

Similar to Lu, the laser ionization of Pr was also significantly enhanced by adding the 532~nm laser with both \textquotedblleft immediate" and \textquotedblleft slow" effects. Unlike Lu, there was an increase of Pr$^+$ by only shooting 532~nm laser into the source. Due to the very complex atomic level structure of Pr, it is difficult to tell if it is the result of a thermal effect or an optical resonance. Another interesting observation is that PrO$^+$ molecular ions ($IP$$_{PrO}$=4.9~eV [\citenum{PrO}])were generated only using the 532~nm laser. Beside PrO$^+$, the molecular ions TiO$^+$ ($IP$$_{TiO}$=6.8~eV [\citenum{TiO}]) and ZrO$^+$ ($IP$$_{ZrO}$=6.2~eV [\citenum{ZrO}])were also observed from the Ti and Zr foil that were used as carriers for the AAS solution. The dependence of these molecular ion yields on the 532~nm laser power was studied (Fig.~\ref{molecular}). To avoid the influence of plasma and thermal effect caused by the powerful 532~nm laser, IGLIS was set in the suppression mode: the repeller had +3~V relative to the source. A small amount of surface ionized Ti$^+$ was also observed in this mode due to the low suppression potential on the repeller relative to the big amount of Ti sample ($\sim$1$\times$1~cm$^2$ 0.001 inch thickness Ti foil) in the crucible. The parabolic dependence of PrO$^+$ and TiO$^+$ on the laser power (linear in logarithmic scale in Fig.~\ref{molecular}) indicates two-photon excitation with nonresonant ionization by the 532~nm laser~[\citenum{Bloembergen}]. The Pr$^+$ and Ti$^+$ ions, which were from the resonant one-photon excitation and the surface ionization respectively, did not show the dependence on the 532~nm laser.

\section{Summary}\label{summary}

Laser resonant ionization scheme developments of lanthanide elements Lu and Pr were performed at the off-line LIS stand of ISAC-TRIUMF. The relative laser ionization efficiencies of different laser schemes have been evaluated. Additionally, on-line tests of Lu schemes have been performed on two different types of targets (uranium carbide and tantalum) and two different types of ion sources (IGLIS and hot cavity RILIS). The temperature dependence of the laser resonance ionization efficiency via Rydberg states of lutetium has been studied. The validity to use Rydberg state schemes on IGLIS was investigated. The ionization mechanism of Rydberg states, especially perturbed ones, was discussed. The generation of molecular ions PrO$^+$, TiO$^+$ and ZrO$^+$ by interaction with 6.7~W 532~nm laser light was observed. Their two-photo excitation character, a linear dependency on the square of the laser intensity, has been measured and discussed.

\section*{Acknowledgements}

The experimental work is funded by TRIUMF which receives federal funding via a contribution agreement with the National Research Council of Canada and through a Natural Sciences and Engineering Research Council of Canada (NSERC) Discovery Grant (386343-2011). M. Mostamand acknowledges funding through the University of Manitoba graduate fellowship.





\bibliographystyle{elsarticle-num}

\begin{thebibliography}{35}


\bibitem{Mumpower} M. R. Mumpower, G. C. McLaughlin and R. Surman, The Rare Earth Peak : An Overlooked r-Process Diagnostic,  The Astrophys. J. 752 (2012) 117-127.

\bibitem{Nayak99} D. Nayak and S. Lahiri, Application of radioisotopes in the field of nuclear medicine, J. Radioanal. Nucl. Chem. 242(1999) 423-432.

\bibitem{Seb14} S. Raeder, H. Heggen, J. Lassen, F. Ames, D. Bishop, P. Bricault, P. Kunz, A. Mjøs and A. Teigelh\"ofer, An ion guide laser ion source for isobar-suppressed rare isotope beams, Rev. Sci. Instrum. 85 (2014) 033309.

\bibitem{Fink15} D. A. Fink \textit{et al.}, On-line implementation and first operation of the Laser Ion Source and Trap at ISOLDE/CERN, Nucl. Instrum. Methods Phys. Res., Sect. B 344 (2015) 83-95.

\bibitem{Meg30} W. Meggers and B. Scribner, Regularities in  the spectra of lutetium, J. Res. Nat. Bur. Stand. 5 (1930) 73-81.

\bibitem{King31} A. King, Temperature Classification of the Spectra of Ytterbium and Lutetium, Astrophys. J. 74 (1931) 328-341.

\bibitem{Klin54} P. Klinkenberg, Analysis of the arc spectrum of lutetium, Physica 21 (1954) 53-62.

\bibitem{Cam72} P. Camus and F. Tomkins, Absorption-line series in Lu I, J. Phys. France 33 (1972) 197-201.

\bibitem{Maeda89} H. Maeda, Y. Mizugai, Y. Matsumoto, A. Suzuki and M. Takami, Highly excited even Rydberg series of Lu I studied by two-step laser photoionisation spectroscopy, J. Phys. B 22 (1989) L511-L516.

\bibitem{Vid92} E. Vidolova-Angelova, D. Angelov, T. Krustev and S. Mincheva, Investigation of lutetium Rydberg states by laser multistep resonance ionization spectroscopy, Z. Phys. D 23 (1992) 215-218.

\bibitem{Miller82} C. Miller and N. Nogar, Autoionizing and high-lying Rydberg states of lutetium atoms, AIP Conf Proc. 90 (1982) 90-94.

\bibitem{Xu93} C. Xu, X. Xu, H. Ma, L. Li, W. Huang, D. Chen and F. Zhu, The study of autoionizing states of lutetium atoms by resonance ionization spectroscopy, J. Phys. B 26 (1993) 2827-2835.

\bibitem{Kujirai98} O. Kujirai and Y. Ogawa, observation of even-parity autoionization states of lutetium atom by optogalvanic spectroscopy, J. Phys. Soc. Jpn. 67 (1998) 1056-1057.

\bibitem{Ogawa99} Y. Ogawa and O. Kujirai, Study of even-parity autoionization states of lutetium atom by atomic beam laser resonance ionization spectroscopy, J. Phys. Soc. Jpn. 68 (1999) 428-433.

\bibitem{Bekov81} G. I. Bekov and E. P. Vidolova-Angelova, Optimal scheme for multistage photoionization of lutetium atoms by laser radiation, Sov. J. Quantum Electron. 11 (1981) 137-139.

\bibitem{Krustev93} T. B. Krustev, S. T. Mincheva, D. A. Angelov and E. P. Vidolova-Angelova, Determination of traces of lutetium in geological samples by resonance ionization spectroscopy, J. Anal. At. Spectrom. 8 (1993) 1029-1031.

\bibitem{NIST} A. Kramida, Yu. Ralchenko, J. Reader and {NIST ASD Team}, {NIST Atomic Spectra Database (ver. 5.3)}, National Institute of Standards and Technology, Gaithersburg, MD, \url{http://physics.nist.gov/asd} (2015).

\bibitem{Meggers42} W. F. Meggers, Atomic Spectra of Rare Earth Elements, Rev. Mod. Phys. 14 (1942) 96-103.

\bibitem{Martin78} W. C. Martin, R. Zalubas and L. Hagan, Atomic Energy Levels - The Rare-Earth Elements, Nat. Stand. Ref. Data Ser. NSRDS-NBS 60 , Nat. Bur. Stand. U.S. 1978.

\bibitem{Kujirai00} O. Kujirai and Y. Ogawa, Study of Odd-Parity Autoionization States of Praseodymium Atom by Optogalvanic Spectroscopy, J. Phys. Soc. Jpn. 69 (2000) 2845-2849.

\bibitem{Ivarsson01} S. Ivarsson, U. Litz\'{e}n and G. M. Wahlgren, Accurate wavelengths, oscillator strengths and hyperfine structure in selected praseodymium lines of astrophysical interest, Phys. Scripta 64 (2001) 455-461.

\bibitem{Iimura95} H. Iimura, M. Kubota, Y. Nakahara, S. Ichikawa and T. Horiguchi, Measurement of nuclear moments of radioactive nuclide, $^{143}$Pr (T$_{1/2}$ = 13.6 d), by collinear laser-ion-beam spectroscopy, proceedings of the 16th international symposium on advanced nuclear energy research, Japan, 1995, 350-357.

\bibitem{Froemmgen13} N. Froemmgen, Collinear laser spectroscopy on radioactive praseodymium ions and cadmium ions, Ph.D. dissertation, Johannes Gutenberg-Universit\"{a}t, Mainz (Germany) 2013.

\bibitem{Wendt14} K. Wendt , T. Gottwald, C. Mattolat and S. Raeder, Ionization potentials of the lanthanides and actinides towards atomic spectroscopy of super-heavy elements, Hyperfine Interact. 227 (2014) 55-67.

\bibitem{Fedosseev12} V. N. Fedosseev \textit{et al.}, Upgrade of the resonance ionization laser ion source at ISOLDE on-line isotope separation facility: New lasers and new ion beams, Rev. Sci. Instrum. 83 (2012) 02A903.

\bibitem{Lav13} J. P. Lavoie, R. Li, P. Bricault, J. Lassen, O. Chachkova, and A. Teigelh\"{o}fer, A test stand for off-line ion source development at TRIUMF, Rev. Sci. Instrum. 84 (2013) 013306.

\bibitem{Li13} R. Li, J. Lassen, A. Teigelh\"ofer, J. P. Lavoie, P. Bricault, O. Chackakova, J. Meissner and Y. Zlateva, Resonant ionization laser ion source {(RILIS)} off-line developments on {Ga}, {Al} and {Ca}, Nucl. Instrum. Methods Phys. Res., B 308 (2013) 74-79.

\bibitem{Wendt12} K. Wendt, C. Geppert, C. Mattolat, G. Passler, S. Raeder, F. Schwellnus, K. Wies and N. Trautmann, Progress of ultra trace determination of technetium using laser resonance ionization mass spectrometry, Anal. and Bioanal. Chem. 404 (2012) 2173-2176.

\bibitem{Tei10} A. Teigelh\"{o}fer, P. Bricault, O. Chachkova, M. Gillner, J. Lassen, J. P. Lavoie, R. Li, J. Meissner, W. Neu, K. Wendt, Grating tuned Ti:Sa laser for in-source spectroscopy of Rydberg and autoionizing states, Hyperfine Interact. 196 (2010) 161-168.

\bibitem{LiLu16} R. Li, J. Lassen, Z. P. Zhong, F. D. Jia, M. Mostamand, X. K. Li, B. B. Reich, A. Teigelh\"{o}fer and E. Yan, Study of even-parity Rydberg and autoionizing states of lutetium by laser resonance ionization spectroscopy, Phys. Rev. A, in Press.

\bibitem{Kunz} P. Kunz, C. Andreoiu, P. Bricault, M. Dombsky, J. Lassen, A. Teigelh\"{o}fer, H. Heggen and F. Wong, Nuclear and in-source laser spectroscopy with the ISAC yield station, Rev. Sci. Instrum. 85 (2014), 53305.

\bibitem{Li16} R. Li, J. Lassen, J. Ruczkowski, A. Teigelh\"{o}fer and P. Bricault, Laser resonance ionization spectroscopy of antimony, Spectrochim. Acta B. 128 (2017), 36-44.

\bibitem{Bransden83} B. H. Bransden and C. J. Joachain, Physics of Atoms and Molecules, Longman, London, 1983.

\bibitem{SIMION} SIMION, Scientific Instrument Services, \url{http:/simion.com/}.

\bibitem{Sandner86} W. Sandner, K. A. Safinya and T. F. Gallagher, Forced autoionization, Phys. Rev. A 33 (1986) 1008-1019.

\bibitem{PrO} R. J. Ackermann, E. G. Rauh and R. J. Thorn, The thermodynamics of ionization of gaseous oxides; the first ionization potentials of the lanthanide metals and monoxides, J. Chem. Phys. 65 (1976) 1027-1031.

\bibitem {TiO} A. D. Sappey, G. Eiden, J. E. Harrington and J. C. Weisshaar, Vibronic structure of TiO$^+$ from multiphoton ionization photoelectron spectroscopy, J. Chem. Phys. 90 (1989) 1415-1428.

\bibitem{ZrO} E. G. Rauh and R. J. Ackermann, First ionization potentials of some refractory oxide vapors, J. Chem. Phys. 60 (1974) 1396-1400.

\bibitem{Bloembergen} N. Bloembergen and M.D. Levenson, High Resolution Laser Spectroscopy, Top. Appl. Phys. 13 (Springer Verlag, Berlin, 1976) P. 315.


\end{thebibliography}



\end{document}